\documentstyle[twoside,fleqn]{article}
\makeatletter
\def\fileversion{v2.6}
\def\filedate{24 November 1993}

\newdimen\@bls                    % \@b(ase)l(ine)s(kip)
\@bls=\baselineskip               % \@bls ~= \baselineskip for \normalsize
\advance\@bls -1ex                % (fudge term)
\newdimen\@eps                    %
\def\section{\@startsection{section}{1}{\z@}
  {1.5\@bls plus 0.5\@bls}{1\@bls}{\normalsize\bf}}
\def\subsection{\@startsection{subsection}{2}{\z@}
  {1\@bls plus 0.25\@bls}{\@eps}{\normalsize\bf}}
\def\subsubsection{\@startsection{subsubsection}{3}{\z@}
  {1\@bls plus 0.25\@bls}{\@eps}{\normalsize\bf}}
\def\paragraph{\@startsection{paragraph}{4}{\parindent}
  {1\@bls plus 0.25\@bls}{0.5em}{\normalsize\bf}}
\def\subparagraph{\@startsection{subparagraph}{4}{\parindent}
  {1\@bls plus 0.25\@bls}{0.5em}{\normalsize\bf}}

\def\@sect#1#2#3#4#5#6[#7]#8{\ifnum #2>\c@secnumdepth
  \def\@svsec{}\else
  \refstepcounter{#1}\edef\@svsec{\csname the#1\endcsname.\hskip0.5em}\fi
  \@tempskipa #5\relax
  \ifdim \@tempskipa>\z@
    \begingroup
      #6\relax
      \@hangfrom{\hskip #3\relax\@svsec}{\interlinepenalty \@M #8\par}%
    \endgroup
    \csname #1mark\endcsname{#7}\addcontentsline
      {toc}{#1}{\ifnum #2>\c@secnumdepth \else
        \protect\numberline{\csname the#1\endcsname}\fi #7}%
  \else
    \def\@svsechd{#6\hskip #3\@svsec #8\csname #1mark\endcsname
      {#7}\addcontentsline{toc}{#1}{\ifnum #2>\c@secnumdepth \else
        \protect\numberline{\csname the#1\endcsname}\fi #7}}%
  \fi \@xsect{#5}}

\long\def\@makefigurecaption#1#2{\vskip 10mm #1. #2\par}

\long\def\@maketablecaption#1#2{\hbox to \hsize{\parbox[t]{\hsize}
  {#1 \\ #2}}\vskip 0.3ex}

\def\fnum@figure{Figure \thefigure}
\def\figure{\let\@makecaption\@makefigurecaption \@float{figure}}
\@namedef{figure*}{\let\@makecaption\@makefigurecaption \@dblfloat{figure}}

\def\table{\let\@makecaption\@maketablecaption \@float{table}}
\@namedef{table*}{\let\@makecaption\@maketablecaption \@dblfloat{table}}

\textfloatsep=\floatsep           % Space between main text and floats
\intextsep=\floatsep              % Space between in-text figures and
\long\def\@makefntext#1{\parindent 1em\noindent\hbox{${}^{\@thefnmark}$}#1}

\def\maketitle{\begingroup        % Initialize generation of front-matter
    \def\thefootnote{\fnsymbol{footnote}}%
    \newpage \global\@topnum\z@
    \@maketitle \@thanks
  \endgroup
  \let\maketitle\relax \let\@maketitle\relax
  \gdef\@thanks{}\let\thanks\relax
  \gdef\@address{}\gdef\@author{}\gdef\@title{}\let\address\relax}

\def\justify@on{\let\\=\@normalcr
  \leftskip\z@ \@rightskip\z@ \rightskip\@rightskip}

\newbox\fm@box                    % Box to capture front-matter in

\def\@maketitle{%                 % Actual formatting of \maketitle
  \global\setbox\fm@box=\vbox\bgroup
    \vskip 8mm                    % 930715: 8mm white space above title
    \raggedright                  % Front-matter text is ragged right
    \hyphenpenalty\@M             % and is not hyphenated.
    {\Large \@title \par}         % Title set in larger font.
    \vskip\@bls                   % One line of vertical space after title.
    {\normalsize                  % each author set in the normal
     \@author \par}               % typeface size
    \vskip\@bls                   % One line of vertical space after author(s).
    \@address                     % all addresses
  \egroup
  \twocolumn[%                    % Front-matter text is over 2 columns.
    \unvbox\fm@box                % Unwrap contents of front-matter box
    \vskip\@bls                   % add 1 line of vertical space,
    \unvbox\abstract@box          % unwrap contents of abstract boxes,
    \vskip 2pc]}                  % and add 2pc of vertical space

\newcounter{address}
\def\theaddress{\alph{address}}
\def\@makeadmark#1{\hbox{$^{\rm #1}$}}

\def\address#1{\addressmark\begingroup
  \xdef\@tempa{\theaddress}\let\\=\relax
  \def\protect{\noexpand\protect\noexpand}\xdef\@address{\@address
  \protect\addresstext{\@tempa}{#1}}\endgroup}
\def\@address{}

\def\addressmark{\stepcounter{address}%
  \xdef\@tempb{\theaddress}\@makeadmark{\@tempb}}

\def\addresstext#1#2{\leavevmode \begingroup
  \raggedright \hyphenpenalty\@M \@makeadmark{#1}#2\par \endgroup
  \vskip\@bls}

\newbox\abstract@box              % Box to capture abstract in

\def\abstract{%
  \global\setbox\abstract@box=\vbox\bgroup
  \small\rm
  \ignorespaces}
\def\endabstract{\par \egroup}

\def\thebibliography#1{\section*{REFERENCES}\list{\arabic{enumi}.}
  {\settowidth\labelwidth{#1.}\leftmargin=1.67em
   \labelsep\leftmargin \advance\labelsep-\labelwidth
   \itemsep\z@ \parsep\z@
   \usecounter{enumi}}\def\makelabel##1{\rlap{##1}\hss}%
   \def\newblock{\hskip 0.11em plus 0.33em minus -0.07em}
   \sloppy \clubpenalty=4000 \widowpenalty=4000 \sfcode`\.=1000\relax}

\newcount\@tempcntc
\def\@citex[#1]#2{\if@filesw\immediate\write\@auxout{\string\citation{#2}}\fi
  \@tempcnta\z@\@tempcntb\m@ne\def\@citea{}\@cite{\@for\@citeb:=#2\do
    {\@ifundefined
       {b@\@citeb}{\@citeo\@tempcntb\m@ne\@citea
        \def\@citea{,\penalty\@m\ }{\bf ?}\@warning
       {Citation `\@citeb' on page \thepage \space undefined}}%
    {\setbox\z@\hbox{\global\@tempcntc0\csname b@\@citeb\endcsname\relax}%
     \ifnum\@tempcntc=\z@ \@citeo\@tempcntb\m@ne
       \@citea\def\@citea{,\penalty\@m}
       \hbox{\csname b@\@citeb\endcsname}%
     \else
      \advance\@tempcntb\@ne
      \ifnum\@tempcntb=\@tempcntc
      \else\advance\@tempcntb\m@ne\@citeo
      \@tempcnta\@tempcntc\@tempcntb\@tempcntc\fi\fi}}\@citeo}{#1}}

\def\@citeo{\ifnum\@tempcnta>\@tempcntb\else\@citea
  \def\@citea{,\penalty\@m}%
  \ifnum\@tempcnta=\@tempcntb\the\@tempcnta\else
   {\advance\@tempcnta\@ne\ifnum\@tempcnta=\@tempcntb \else
\def\@citea{--}\fi
    \advance\@tempcnta\m@ne\the\@tempcnta\@citea\the\@tempcntb}\fi\fi}

\def\ps@crcplain{\let\@mkboth\@gobbletwo
     \def\@oddhead{\reset@font{\sl\rightmark}\hfil \rm\thepage}%
     \def\@evenhead{\reset@font\rm \thepage\hfil\sl\leftmark}%
     \let\@oddfoot\@empty
     \let\@evenfoot\@oddfoot}

\ps@crcplain                    % modified 'plain' page style

\newcommand{\AmS}{{\protect\the\textfont2
  A\kern-.1667em\lower.5ex\hbox{M}\kern-.125emS}}

\makeatother

\title{Spectral Zeta Functions in Non-Commutative
Spacetimes}

\author{{Emilio Elizalde}
\address{Instituto de Ciencias del Espacio (CSIC) \& IEEC, Edifici Nexus,
Gran Capit\`{a} 2-4, 08034 Barcelona }
\address{Departament ECM i IFAE,
Facultat de F\'{\i}sica,  Universitat de Barcelona, Diagonal 647,
08028 Barcelona, Spain}%
\thanks{Invited contribution to the {\sl International Conference
on Quantum Gravity and Spectral Geometry}, Neapel, Italy, July
1--7, 2001.}
\thanks{Supported by DGI/SGPI (Spain), project
BFM2000-0810, by CIRIT (Generalitat de Catalunya),
contract 1999SGR-00257, and by the program INFN (Italy)--DGICYT (Spain).}}

\begin{document}
\begin{abstract}
Formulas for the most  general case of the zeta function associated to a
quadratic+linear+constant form (in {\bf Z}) are given. As examples,
the spectral zeta functions $\zeta_\alpha (s)$ corresponding to
bosonic ($\alpha =2$) and to fermionic ($\alpha =3$) quantum fields living
on a noncommutative, partially toroidal spacetime are investigated.
Simple poles show up at $s=0$, as well as in other places (simple or double,
depending on the
number of compactified, noncompactified, and noncommutative dimensions of the
spacetime). This poses a challenge
to the zeta-function regularization procedure.
\end{abstract}

\maketitle

A fundamental property shared by all zeta functions is the
existence of a reflection formula. It allows for its analytic continuation
in an easy way \cite{zb1}. But much better than that is a
formula that yields {\it exponentially quick convergence} and {\it  everywhere}, not
just in the reflected domain of convergence.
 We here provide explicit, Chowla-Selberg-like \cite{cs}
extended formulas for {\it all} possible cases involving forms of the
very general type: quadratic+linear+constant \cite{eli2a}-\cite{eejpa01}.
Then we move to specific applications of these formulas
in non-commutative field theory.

Consider the zeta function (Re $s > p/2$):
\begin{eqnarray}
\hspace*{-1mm} \zeta_{A,\vec{c},q} (s)& =& {\sum_{\vec{n} \in \mbox{\bf Z}^p}}'
 \left[
\frac{1}{2}\left( \vec{n}+\vec{c}\right)^T A
\left( \vec{n}+\vec{c}\right)+ q\right]^{-s} \nonumber \\  &\equiv&
{ \sum_{\vec{n} \in \mbox{\bf Z}^p}}'
\left[ Q\left( \vec{n}+\vec{c}\right)+ q\right]^{-s}.  \label{zf1}
\end{eqnarray}
The prime on a summation sign  means
that the point $\vec{n}=\vec{0}$ is to be excluded from the sum.
This is irrelevant when $q$ or some component of $\vec{c}$
is non-zero but, on the contrary, it becomes an inescapable condition
in the case when $c_1=\cdots =c_p=q=0$. We can
view the expression inside the square brackets of the zeta function as a
sum of a quadratic, a linear, and a constant form, namely,
$Q\left( \vec{n}+\vec{c}\right)+ q = Q( \vec{n}) + L(\vec{n})+ \bar{q}$.

(i) \, For $q\neq 0$, we obtain
\cite{eecmp1}
\begin{eqnarray}
  \zeta_{A,\vec{c},q} (s)& =& \frac{(2\pi )^{p/2} q^{p/2 -s}}{
\sqrt{\det A}} \, \frac{\Gamma(s-p/2)}{\Gamma (s)} \nonumber \\  &+&
 \frac{2^{s/2+p/4+2}\pi^s q^{-s/2 +p/4}}{\sqrt{\det A} \
\Gamma (s)}  \label{qpd1}\\ && \hspace*{-12mm}\times {\sum_{\vec{m} \in
\mbox{\bf Z}^p_{1/2}}}' \cos
(2\pi
 \vec{m}\cdot \vec{c}) \left( \vec{m}^T A^{-1} \vec{m}
\right)^{s/2-p/4}  \nonumber \\  &\times& K_{p/2-s} \left( 2\pi \sqrt{2q \,
 \vec{m}^T A^{-1} \vec{m}}\right), \nonumber
\end{eqnarray}
where $K_\nu$ is the modified Bessel function of the second kind and
the subindex 1/2 in
$\mbox{\bf Z}^p_{1/2}$ means that in this sum, only half of the vectors
$\vec{m} \in \mbox{\bf Z}^p$ enter.
We have
denoted this formula, Eq. (\ref{qpd1}), by the acronym ECS1.
It is notorious  how the only pole of this inhomogeneous Epstein
zeta function appears, explicitly, at $s=p/2$, where it belongs. Its
residue is given by:
\begin{eqnarray}
\mbox{Res}_{s=p/2}  \zeta_{A,\vec{c},q} (s) =
\frac{(2\pi )^{p/2}}{\Gamma(p/2)}\, (\det A)^{-1/2}.
\end{eqnarray}
\medskip

(ii) \, In the case  $q=0$ but $c_1\neq 0$, we obtain \cite{eejpa01}:
 \begin{eqnarray}
 && \hspace*{-6mm} \zeta_{A_p,\vec{c},0} (s) = \frac{2^{s}}{\Gamma (s)} \,
 \left(\det{A_{p-1}} \right)^{-1/2} \left\{  \pi^{(p-1)/2} \left( a_{11}
\right. \right. \nonumber \\ &&- \left.
\vec{a}_{p-1}^T A_{p-1}^{-1}\vec{a}_{p-1} \right)^{(p-1)/2-s}
\Gamma \left( s-(p-1)/2 \right)  \nonumber
\\  && \times
\left[ \zeta_H (2s-p+1,c_1) + \zeta_H (2s-p+1,1-c_1)\right] \nonumber
\\  &&+ 4
\pi^s \left( a_{11}-\vec{a}_{p-1}^T
A_{p-1}^{-1}\vec{a}_{p-1} \right)^{(p-1)/4-s/2} \nonumber
\\  &&  \hspace*{-12mm} \times  \sum_{n_1 \in \mbox{\bf Z}}
{\sum_{\vec{m} \in \mbox{\bf Z}^{p-1}_{1/2}}}' \cos \left[ 2\pi
 \vec{m}^T \left( \vec{c}_{p-1}+ A_{p-1}^{-1}\vec{a}_{p-1}(n_1+c_1 )
\right)\right] \nonumber
\\  & \times& |n_1+c_1|^{(p-1)/2-s} \left( \vec{m}^TA_{p-1}^{-1}
\vec{m}\right)^{s/2-(p-1)/4} \nonumber
\\  &&  \hspace*{-24mm}\left. K_{(p-1)/2-s}\left(2\pi |n_1+c_1| \sqrt{
\left( a_{11}-\vec{a}_{p-1}^T A_{p-1}^{-1}\vec{a}_{p-1} \right)
\vec{m}^TA_{p-1}^{-1}\vec{m}}\right)\right\} \nonumber
\\  &-& \left( \frac{1}{2}
\vec{c}^TA\vec{c}
\right)^{-s}.
 \label{ecsc1}
\end{eqnarray}
 $A_{p-1}$ is  the submatrix of $A_p$ composed of the last
$p-1$ rows and columns. Moreover, $a_{11}$ is  the first diagonal
component of $A_p$,
while $\vec{a}_{p-1} =(a_{12}, \ldots, a_{1p})^T =(a_{21}, \ldots, a_{p1})^T$,
and $\vec{m}=(n_2,\ldots, n_p)^T$. Note that this is an {\it explicit formula},
that the only pole at $s=p/2$ appears also explicitly, and that the second term
of the rhs is a series of exponentially fast convergence. It has,
therefore (as Eq. (\ref{qpd1})), all the properties required to qualify as
a CS formula. We  name this expression ECS2.

The diagonal subcase is rather more simple:
 \begin{eqnarray}
&& \hspace*{-9mm}    \zeta_{A_p,\vec{c},0} (s) = \frac{2^{s}}{\Gamma (s)} \,
 \left(\det{A_{p-1}} \right)^{-1/2} \left\{  \pi^{(p-1)/2}  a_{1}^{(p-1)/2-s}
\right. \nonumber
\\  &&  \times \,\Gamma \left( s-(p-1)/2 \right) \,
\left[ \zeta_H (2s-p+1,c_1)\right. \nonumber
\\  &&  \left. + \zeta_H (2s-p+1,1-c_1)\right]  \nonumber
\\  &&  \hspace*{-3mm} + 4 \pi^s a_{1}^{(p-1)/4-s/2} \sum_{n_1 \in
\mbox{\bf Z}}
{\sum_{\vec{m} \in \mbox{\bf Z}^{p-1}_{1/2}}}' \cos \left( 2\pi
 \vec{m}^T  \vec{c}_{p-1}\right) \nonumber
\\  && \times |n_1+c_1|^{(p-1)/2-s}  \left(
\vec{m}^TA_{p-1}^{-1}\vec{m}\right)^{s/2-(p-1)/4}\nonumber
\\  && \times \, \left. K_{(p-1)/2-s}\left(2\pi |n_1+c_1| \sqrt{
 a_{1}\vec{m}^TA_{p-1}^{-1}\vec{m}}\right)\right\}\nonumber
\\  &&  - \left( \frac{1}{2}
 \vec{c}^TA\vec{c} \right)^{-s}.
\label{ecsc2}
\end{eqnarray}
We call this formula ECS2d.

(iii) \, In the case  $c_1=\cdots =c_p=q=0$ we have similar expressions. Let us
just write down the one corresponding to the diagonal case:
 \begin{eqnarray}
&& \hspace*{-6mm}  \zeta_{A_p} (s)  =
\frac{2^{1+s}}{\Gamma (s)} \,
\sum_{j=0}^{p-1} \left(\det{A_j} \right)^{-1/2} \left[ \pi^{j/2}
a_{p-j}^{j/2-s} \right.
\nonumber \\  && \times \Gamma \left( s-j/2 \right) \, \zeta_R(2s-j)
 + 4 \pi^s a_{p-j}^{j/4-s/2}\nonumber \\  && \times
\sum_{n=1}^\infty
{\sum_{\vec{m}_j \in \mbox{\bf Z}^j}}' n^{j/2-s}
\left(\vec{m}_j^t A_j^{-1}\vec{m}_j\right)^{s/2-j/4}\nonumber \\  && \times
\left.
K_{j/2-s}\left(2\pi n \sqrt{ a_{p-j} \vec{m}_j^t A_j^{-1}\vec{m}_j}\right)
\right],\label{ecs1}
\end{eqnarray}
We call this formula ECS3d.
This formula, Eq. (\ref{ecs1}),  provides a convenient
analytic continuation of the zeta function to the whole complex
plane, with its only simple pole showing up explicitly. Aside from
this, the finite part of the first sum in the expression is quite
easy to obtain, and the remainder ---an awfully looking multiple
series--- is in fact an extremely-fast convergent sum, yielding a
 small contribution, as happens in the
CS formula. In fact, since it corresponds to the
case $q=0$, this expression should be viewed as {\it the}
extension of the original  Chowla-Selberg formula ---for the zeta
function associated with an homogeneous quadratic form in two
dimensions--- to an arbitrary number, $p$, of dimensions. The rest
of the formulas above provide also extensions of the original CS expression.    The investigation of the
general case of a quadratic+linear+constant form has been thus completed.

We shall now consider the physical example of a quantum system consisting
of scalars and vector fields on a
$D-$dimensional noncommutative manifold, $M$, of the form
$\mbox{\bf R}^{1,d}\bigotimes \mbox{\bf T}_\theta^p$ (thus $D=d+p+1$).
$\mbox{\bf T}_\theta^p$ is a $p-$dimensional noncommutative torus,
its coordinates satisfying the usual relation: $[x_j,x_k]=i\theta
\sigma_{jk}$. Here $\sigma_{jk}$ is a real nonsingular, antisymmetric
matrix of $\pm 1$ entries, and $\theta$ is
the noncommutative parameter.

This physical system has attracted much interest.
It has been  shown, in particular, that
noncommutative gauge theories describe the low energy excitations
of open strings on $D-$branes in a background Neveu-Schwarz
two-form field \cite{connes,douglas,seiberg}.
 This interesting
system provides us with a quite non-trivial case where the
formulas derived above are indeed useful. For one, the zeta
functions corresponding to bosonic and fermionic operators in this
system are of a different kind, never considered before. And, moreover,
they can be most conveniently written in terms of the zeta functions
above. What is also nice is the fact that a unified
treatment (with just {\it one} zeta function) can be given for both cases,
the nature of the field appearing there as a simple parameter, together with
those corresponding to the numbers of compactified, noncompactified,
and noncommutative dimensions of the spacetime.

The spectral zeta function for the corresponding
(pseudo-)differential operator can be written in the form \cite{bgz}
\begin{eqnarray}
&& \hspace*{-4mm} \zeta_{\alpha} (s) = \frac{V}{(4\pi )^{(d+1)/2}}
\frac{\Gamma (s-(d+1)/2)}{\Gamma (s)} \label{za1} \\
&& \hspace*{-8mm} \times {\sum_{\vec{n} \in \mbox{\bf
Z}^p}}' Q (\vec{n})^{(d+1)/2-s} \left[ 1+ \Lambda
\theta^{2-2\alpha}Q (\vec{n})^{-\alpha}\right]^{(d+1)/2-s},
\nonumber
\end{eqnarray}
where $V=$ Vol\,($\mbox{\bf R}^{d+1})$, the volume of the
non-compact part, and $Q (\vec{n}) = \sum_{j=1}^p a_j n_j^2$, a
diagonal quadratic form, being the compactification radii
$R_j=a_j^{-1/2}$. Moreover, the value of the parameter $\alpha =2$ for
scalar fields and $\alpha =3$ for vectors, distinguishes between the
two different fields. In the particular case when we set all
the compactification radii equal to $R$, we obtain:
\begin{eqnarray}
&&\hspace*{-4mm} \zeta_{\alpha} (s) = \frac{V}{(4\pi )^{(d+1)/2}}
\frac{\Gamma (s-(d+1)/2)}{\Gamma (s) R^{d+1-2s}} \nonumber \\
&& \hspace*{-1mm} \times
 {\sum_{\vec{n}
\in \mbox{\bf Z}^p}}' I (\vec{n})^{(d+1)/2-s} \left[ 1+ \Lambda
\theta^{2-2\alpha}\right. \nonumber \\
&& \hspace*{-1mm} \left.\times  R^{2\alpha} I
(\vec{n})^{-\alpha}\right]^{(d+1)/2-s}, \label{za11}
\end{eqnarray}
being now the quadratic form: $I (\vec{n}) = \sum_{j=1}^p n_j^2$.

After some calculations, this zeta function can be written in
terms of the zeta functions that I have considered before, with the
result:
\begin{eqnarray}
&&\hspace*{-6mm}\zeta_{\alpha} (s) = \frac{V}{(4\pi )^{(d+1)/2}}
\sum_{l=0}^\infty
\frac{\Gamma (s+l-(d+1)/2)}{l! \, \Gamma (s)}
\nonumber \\ && \hspace*{-4mm} \times  (-\Lambda
\theta^{2-2\alpha})^l \ \zeta_{Q,\vec{0},0}(s+\alpha l-(d+1)/2),
\label{za2}
\end{eqnarray}
which reduces, in the particular case of equal radii, to
\begin{eqnarray}
&&\hspace*{-8mm}\zeta_{\alpha} (s) = \frac{V}{(4\pi )^{(d+1)/2}R^{d+1-2s}}
\sum_{l=0}^\infty \frac{\Gamma (s+l-(d+1)/2)}{l! \, \Gamma (s)}\,
\nonumber \\ && \hspace*{-1mm} \times
(-\Lambda \theta^{2-2\alpha})^l \ \zeta_E(s+\alpha l-(d+1)/2),
\label{za21}
\end{eqnarray}

The pole structure of the resulting zeta function deserves a careful
analysis. It differs, in fact, very much from all cases that were
known in the literature till now. This is not difficult to understand,
from the fact that the pole of the Epstein zeta function at
$s=p/2-\alpha k+(d+1)/2=D/2-\alpha k$, when combined with the poles of the
gamma functions, yields a very rich pattern of singularities for
$\zeta_\alpha (s)$, on taking into account the different possible
values of the
parameters involved.

Having already given the formula (\ref{za2}) above ---that
contains everything needed to perform such calculation of pole
position, residua and finite part--- for its importance for the
calculation of the determinant and the one-loop effective action
from the zeta function,  one start by specifying what happens at
$s=0$.  Remarkably enough, a pole appears in many cases, depending
on the values of the parameters, as observed by Bytsenko, Gon\c
calves and Zerbini \cite{bgz}. This illustrates what one has to
expect for other values of $s$.  The general case will be
considered later. It is convenient to classify all possible
subcases according to the values of $d$ and $D=d+p+1$. The
calculation has been carried out in \cite{eejpa01} in detail, with
the result:
\begin{table}[bht]
\vspace*{-8mm}

\begin{eqnarray}
&&\hspace*{-8mm}\mbox{For } \, d=2k: \nonumber \\ && \nonumber
\\ &&\hspace*{-8mm}
\left\{ \begin{array}{ccc} \mbox{if }
\, D\neq \dot{\overline{2\alpha}} & \Longrightarrow & \zeta_\alpha
(0) =0, \\   \mbox{if } \, D = \dot{\overline{2\alpha}} &
\Longrightarrow & \zeta_\alpha (0) =\mbox{finite}. \end{array}
\right. \nonumber
\end{eqnarray}
\begin{eqnarray}
&&\hspace*{-8mm}\mbox{For } \, d=2k-1: \nonumber \\ &&  \nonumber
\\ &&\hspace*{-14mm}
\left\{ \begin{array}{ccc} \mbox{if }
\, D\neq \dot{\overline{2\alpha}} \  \left\{ \begin{array}{ll}
\mbox{finite}, & \mbox{for } \, l\leq k \\ 0, & \mbox{for } \, l
> k
\end{array} \right\} & \Longrightarrow & \zeta_\alpha (0) =\mbox{finite},
\\ \mbox{} \\ \mbox{if } \, D = 2\alpha l \  \left\{ \begin{array}{ll}
\mbox{pole}, & \mbox{for } \, l\leq k \\  \mbox{finite}, &
\mbox{for } \, l > k
\end{array} \right\}  & \Longrightarrow &
\zeta_\alpha (0) =\mbox{pole}. \end{array} \right. \nonumber
%\label{za3}
\end{eqnarray}
\caption{{\protect\small Pole structure of the zeta function
$\zeta_\alpha (s)$, at $s=0$, according to the different possible
values of $d$ and $D$ ($\dot{\overline{2\alpha}}$ means {\it multiple} of
$2\alpha$.)}}
\vspace*{-5mm}

\end{table}

 Substituting the corresponding formula, from the
preceding section, for the Epstein zeta functions in Eq.
(\ref{za2}), we obtain the following explicit analytic
continuation of $\zeta_\alpha (s)$ ($\alpha =2,3$),
for bosonic and fermionic fields, to the {\it whole} complex
$s-$plane:
\begin{eqnarray}
&&\hspace*{-8mm}
\zeta_{\alpha} (s) = \frac{2^{s-d} \, V}{(2\pi )^{(d+1)/2}\Gamma
(s)} \sum_{l=0}^\infty \frac{\Gamma (s+l-(d+1)/2)}{l! \, \Gamma
(s+\alpha l -(d+1)/2)}\nonumber \\ &&\hspace*{-2mm}\times \,
 (-2^\alpha \Lambda \theta^{2-2\alpha})^l
\sum_{j=0}^{p-1} \left(\det{A_{j}} \right)^{-1/2} \nonumber
\\ && \hspace*{-8mm} \times \,  \left[ \pi^{j/2} a_{p-j}^{-s-\alpha l
+(d+j+1)/2} \Gamma (s+\alpha l-(d+j+1)/2)
 \right. \nonumber \\ && \hspace*{-2mm}\times \,  \zeta_R(2s+2\alpha
l-d-j-1)  \nonumber \\ && \hspace*{-8mm} +  4 \pi^{s+\alpha
l-(d+1)/2} a_{p-j}^{-(s+\alpha l)/2-(d+j+1)/4}
 \nonumber \\ && \hspace*{-8mm}\times \,  \sum_{n=1}^\infty
{\sum_{\vec{m}_j \in \mbox{\bf Z}^j}}' n^{(d+j+1)/2-s-\alpha l}
 \label{za4} \\ &&   \hspace*{-2mm} \times  \left(\vec{m}_j^t
A_j^{-1}\vec{m}_j\right)^{(s+\alpha l)/2-(d+j+1)/4} \nonumber \\ &&
\left. \hspace*{-8mm}\times \,
K_{(d+j+1)/2-s-\alpha l}\left(2\pi n \sqrt{ a_{p-j} \vec{m}_j^t
A_j^{-1}\vec{m}_j}\right) \right]. \nonumber
\end{eqnarray}
The non-spurious poles of this zeta function are to be found in the
terms corresponding to $j=p-1$.
The pole structure is summarized in Table 2.

\begin{table}[htb]
\vspace*{-6mm}

\begin{center}

\begin{tabular}{|c||c|}
\hline \hline  & $D$\,  even   \\
 \hline \hline
 $p$\,  odd & (1a) \ \, {\it pole} / finite ($l\geq l_1$)  \\ \hline
 $p$\,  even & (1b) \ \, {\it double pole} / pole ($l\geq l_1,l_2$)
 \\ \hline
 \hline  & $D$\,  odd   \\ \hline \hline
 $p$\,  odd  & (2a) \ \,
{\it pole} / pole  \\ \hline
 $p$\,  even  & (2b) \
 \,
{\it pole} / double pole ($l\geq l_2$) \\ \hline \hline

\end{tabular}

\vspace*{3mm}

\caption{{\protect\small General pole structure of the zeta
function $\zeta_\alpha (s)$, according to the different possible
values of $D$ and $p$ being odd or even. In italics, the type of behavior
corresponding to lower values of $l$ is quoted, while the behavior shown
in roman characters corresponds to larger values of $l$. }}

\end{center}
\vspace*{-7mm}

\end{table}
Depending on $D$ and $p$ being even or odd, completely
different situations arise, for different values of $l$: from
the disappearance of the pole, giving rise to a finite
contribution, to the appearance of a simple or a double pole.

An application of these formulas to the calculation of the
one-loop partition function corresponding to quantum fields at finite
temperature, on a noncommutative flat spacetime, is given elsewhere
\cite{bez1}.

\bigskip

Thanks are extended to Andrei Bytsenko and to Sergio Zerbini for stimulating
discussions and  collaboration on this subject.

\end{document}